\documentclass[12pt]{article}
\usepackage[usenames]{color}
\usepackage{setspace}
\usepackage{graphicx}
\usepackage{amsmath}
\usepackage{amssymb}
\usepackage{amsthm}
\usepackage{multirow}
\numberwithin{equation}{section}
\newcommand{\del}{\partial}

\newcommand{\bequ}{\begin{align}}
\newcommand{\eequ}{\end{align}}
\newcommand{\beqn}{\begin{align}}
\newcommand{\eeqn}{\end{align}}
\newcommand{\bctr}{\begin{center}}
\newcommand{\ectr}{\end{center}}
\newcommand{\bit}{\begin{itemize}}
\newcommand{\eit}{\end{itemize}}

\newcommand{\half}{{\frac12}}
\def\e{{\textrm e}}
\def\del{\partial}
\def\half{{\frac12}}

\def\del{\partial}
\def\half{{\frac12}}

\def\del{\partial}
\def\dslash{\del\kern-0.55em\raise 0.14ex\hbox{/}}

\def\rough#1{\raise.3ex\hbox{$#1$\kern-.75em\lower1ex\hbox{$\sim$}}}
\newcommand{\PRD}[3]{{\it Phys. Rev.} {\bf D{#1}} (19{#3}) {#2}}

\newcommand{\PRDM}[3]{{\it Phys. Rev.} {\bf D{#1}} {#2} (20{#3})}

\newcommand{\PLB}[3]{{\it Phys. Lett.} {\bf B{#1}} (19{#2}) {#3}}

\newcommand{\PTP}[3]{{\it Prog. Theor. Phys.} {\bf {#1}} (19{#3}) {#2}}
\newcommand{\PTPM}[3]{{\it Prog. Theor. Phys.} {\bf {#1}} (20{#3}) {#2}}

\newcommand{\ANN}[3]{{\it Ann. Phys. (N.Y.)} {\bf {#1}}, {#2} (19{#3})}

\newcommand{\MPL}[3]{{\it Mod. Phys. Lett.} {\bf A{#1}} (19{#3}) {#2}}
\newcommand{\MPLM}[3]{{\it Mod. Phys. Lett.} {\bf A{#1}} (20{#3}) {#2}}

\newcommand{\jhep}[3]{{\it JHEP} {\bf {#1}} (20{#2}) {#3}}

\newcommand{\hepph}[1]{{\tt hep-ph/#1}}
\begin{document}
\begin{flushright}
{\small KOBE-TH-24-01}\\%
%[-1mm] hep-ph/yymmnnn%
\end{flushright}
\begin{center}
{\LARGE\bf 
Mode Recombination Formula \\and \\
Nonanalytic Term in Effective Potential\\
at Finite Temperature on Compactified Space \\
}
\vskip 1.4cm
Makoto Sakamoto$^{(a)}$
\footnote{E-mail: dragon@kobe-u.ac.jp} and
Kazunori Takenaga$^{(b)}$
\footnote{E-mail: takenaga@kumamoto-hsu.ac.jp}
\\
\vskip 1.0cm
${}^{(a)}$ {\it Department of Physics, Kobe University, 
Rokkodai Nada, Kobe, 657-8501 Japan}
\\[0.2cm]
${}^{(b)}$ {\it Faculty of Health Science, Kumamoto
Health Science University, Izumi-machi, Kita-ku, Kumamoto 861-5598, Japan}
\\
%%%%%
\vskip 1.5cm
%%%
\begin{abstract}
We develop a new formula called a mode recombination formula, and we can
recast the effective potential at finite temperature in one-loop approximation for fermion and scalar fields
on the $D$-dimensional spacetime, $S_{\tau}^1\times R^{D-(p+1)}\times \prod_{i=1}^pS_i^1$
into a convenient form for discussing nonanalytic terms, which cannot be written in the
form of any positive integer power of the field-dependent mass squared, in the effective potential. 
The formula holds irrespective of whether the field is a fermion or a scalar and of boundary conditions 
for spatial $S_i^1$ directions and clarifies the importance of zero modes in the Matsubara and 
Kaluza-Klein modes for the existence of the nonanalytic terms. The effective potential 
is drastically simplified further to obtain the nonanalytic terms in easier and more transparent way. 
In addition to reproducing previous results, we 
find that there exists no nonanalytic term for the fermion field with 
arbitrary boundary condition for the spatial $S_i^1$ direction,  which is also the 
case for the scalar field with the antiperiodic boundary condition for the spatial direction.
\end{abstract}
\end{center}
\vskip 1.0 cm
\newpage
%%%%%%%%%%%%%%%%%%%%%%%%
%
%
%
%
%%%%%%%%%%%%%%%%%%%%%%
\section{Introduction}
%%%%%%%%%%%%%%%%%%%%%%
%
%
%
%
Quantum field theory at finite temperature has provided a fundamental theoretical framework
in high energy physics. In particular, the effective potential at finite temperature
is a crucial tool to investigate physical phenomena involved with the order of the phase transition and its strength.

In the pioneering work by the Dolan-Jackiw\cite{dj}, they found
that there exists a nonanalytic term, which cannot be written in the form of
any positive integer power of field-dependent mass squared, in the effective potential at finite
temperature for a real scalar field. The nonanalytic term obtained by them has three-halves power of the
mass squared, and it turns out to play an essential role to trigger the first order 
phase transition\cite{dine, quiros}. Moreover, the magnitude of the term is related with the strength of the first order phase 
transition, which, in turn, put certain constraint on physical quantity such as the Higgs mass, for example, in the 
study of electroweak baryogenesis\cite{baryon}. Hence, the nonanalytic term in the effective potential 
is an important quantity that must be studied in detail.  

%%%%%%%% NEW %%%%%
Quantum field theory with compactified dimensions has been one of the attractive approaches 
for physics beyond the standard model. Orbifold compactification provides a
framework for gauge-Higgs unification, where the Higgs field is unified into higher 
dimensional gauge fields\cite{HIL,KLY}.
%
%and it can be an alternative solution to the gauge hierarchy problem\cite{HIL,KLY}. 
%
The order of the finite temperature phase transition in the gauge-Higgs unification 
has been studied in \cite{panico,marutake1}, and the first order phase transition actually
occurs due to the 
%
%term with three-halves power of the field-dependent mass squared 
%
nonanalytic term in the effective potential. It has been also shown 
that the quantum field theory at finite temperature with compactified dimensions can possess 
rich phase structures \cite{HOST1, sakatake}. Compactified dimensions also provide the theoretical framework 
for studying quantum field theory itself. For instance, from a point of view of dimensional reduction\cite{cava1, cava2}, models 
with several numbers of $S^1$ have been investigated.
%%%%%%%%%%

In the previous paper\cite{sakatake2022}, we studied all the possible nonanalytic terms in the effective potential at finite temperature 
in one-loop approximation for a real scalar field on the $D$-dimensional 
spacetime, $S_{\tau}^1\times R^{D-(p+1)}\times \prod_{i=1}^pS_i^1$, where the $S_{\tau}^1, R^{D-(p+1)}, S_i^1$ stand for 
the Euclidean time direction, the $D-(p+1)$ dimensional flat Euclidean space, the spatial compactified 
direction, respectively. The effective potential contains the modified Bessel function of the second kind accompanied with 
multiple mode summations with respect to the winding mode associated with each $S^1$. By using the integral representation 
for the modified Bessel function given by the inverse Mellin 
transformation\cite{bromwich, davis} and the analytical extension for the mode summation\cite{elizalde},
we recast the effective potential into the integral form on the 
complex plane\footnote{The studies for dimensional reduction based on the integral form have been carried out in 
Refs.\cite{cava1, cava2}. } and performed the residue integration in order to obtain the nonanalytic terms. We found 
that the only nonanalytic power comes from a term $M^{D-(p+1)}$ which is 
not analytic in $M^2$ when the dimension of the flat Euclidian space, $D-(p+1)$ is odd.
We obtained the coefficient of this term, but other nonanalytic terms proportional to $\log M$ were not studied.
%
%
%there exists only one nonanalytic term when the dimension of the 
%flat Euclidean space, $D-(p+1)$ is odd. 
%

In this paper, we develop a new formula called a mode recombination formula, which plays a  central role for
discussions in the present paper. The effective potential can be recast 
into the convenient form for studying the nonanalytic terms. The formula also clarifies that 
only  the zero mode in the Kaluza-Klein mode associated with each $S^1$ is crucial for the existence of the nonanalytic terms in 
the case of the scalar field satisfying the periodic boundary condition for the spatial $S_i^1~(i=1,\cdots, p)$ direction. 
Then, the effective potential relevant for the nonanalytic terms can be simplified further drastically and is
given in terms of the contribution of the single mode summation with respect to the winding mode
associated with each $S^1$. This is quite different from the previous paper, where it includes the contribution of the 
multiple mode summations. The integral form for the simplified effective potential
on the complex plane is easy to perform the residue integration in order to obtain the nonanalytic terms. 
We reproduce the previous results in easier and more transparent way.

The mode recombination formula holds irrespective of whether the
field is a fermion or a scalar and of the boundary condition for the spatial $S_i^1(i=1,\cdots, p)$ direction. 
The formula also provides a convenient form for studying the nonanalytic terms in the effective potential for the case of the fermion 
field with arbitrary boundary condition for the $S_i^1$ direction. The zero mode for the Euclidean time direction is removed due 
to the antiperiodic boundary condition followed from the quantum statistics for the fermion. This changes the pole structure of 
the analytical extension for the mode summation with respect to the Matsubara mode compared with 
that of the scalar case. We find that there is no nonanalytic term for the case of the fermion. This immediately 
implies that the effective potential does not possess the nonanalytic term for the case of the
real scalar field satisfying the antiperiodic boundary condition for at least one spatial $S_i^1$ direction. 
%
%The result for the case of higher dimensional gauge field is also obtained directly by the result for the scalar filed.
%

This paper is organized as follows. We present the setup in the next section. We prove the mode
recombination formula and present the convenient form of the effective potential for studying the 
nonanalytic terms in each case of the fermion and the scalar in the section $3$. We reproduce the same result
as the one in the previous paper in easier and more transparent way in the section $4$. 
We also study the nonanalytic terms for the case of fermion with arbitrary boundary 
condition in the section $5$ and of the real scalar with the antiperiodic boundary condition in the section $6$. 
The final section is devoted to conclusions and discussions which also include
the case for a higher dimensional gauge field.
%
%
%
%
%
%
%
%
%
%
%%%%%%%%%%%%%%%%%%%%%%%%%%%%%%%%%%%%%%%%%%
\section{Setup}
%%%%%%%%%%%%%%%%%%%%%%%%%%%%%%%%%%%%%%%%%%
%
%
Let us first present the setup for our discussions. We study nonanalytic terms in the effective potential
at finite temperature in one-loop approximation for a real scalar (fermion) field on the $D$-dimensional 
spacetime, $S_{\tau}^1\times R^{D-(p+1)}\times \prod_{i=1}^{p}S^1_i$. We 
employ the Euclidean time formulation for finite temperature quantum field theory and then the
Euclidean time direction, whose coordinate is denoted by $\tau$, is compactified on $S^1_{\tau}$. 
The spatial $p$ directions are compactified on the $p$ numbers of $S^1$ and their coordinates are $y^i~(i=1,\cdots, p)$. 
We denote the circumference of each $S_i^1$ as $L_i~(i=0,1,\cdots, p)$ and $L_0$ stands for the inverse temperature $T^{-1}$.
The $R^{D-(p+1)}$ is the $D-(p+1)$ dimensional flat Euclidean space whose coordinates are $x^k~(k=1,\cdots, D-(p+1))$.

The Lagrangian is given by
 \begin{align}
 {\cal L}&=\half (\del_N \phi)^2 -\frac{m_s^2}{2}\phi^2 -\frac{\lambda}{4!}\phi^4
 +\bar \psi(i\Gamma_N\del_N +m_f)\psi + g\phi\bar\psi\psi\,,
 \label{2.1}
 \end{align}
where the $N$ stands for $N=(\tau, k, i)$, and $\phi\,(\psi)$ is the scalar (fermion) field whose bulk mass is
$m_s\,(m_f)$. The $g$ is the Yukawa coupling.

One needs to specify the boundary conditions for the $S_{\tau}^1$ and $S_i^1 (i=1,\cdots, p)$ directions. For a given field $\Phi(\tau, x^k, y^i)$,
the boundary condition for the $S_{\tau}^1$ direction is specified by
\begin{align}
\Phi(\tau+L_0, x^k, y^i)=\e^{2\pi i\eta_0}\Phi(\tau, x^k, y^i).
 \label{2.2}
\end{align}
The parameter $\eta_0$ is definitely determined by quantum statistics to be $0$ (periodic) for the scalar field or to be
$\half$ (antiperiodic) for the fermion field. On the other hand, the boundary condition for the 
$S_i^1 (i=1,\cdots, p)$ direction is parametrized by
\begin{align}
\Phi(\tau, x^k, y^i+L_i)&=\e^{2\pi i\eta_i}\Phi(\tau, x^k, y^i).
 \label{2.3}
\end{align}
The parameter $\eta_i$ can take $0$ or $\half$ for the real scalar field and can be arbitrary for the fermion field.

We employ the standard prescription to calculate the effective potential at finite temperature in 
one-loop approximation. Let us quickly review the calculations given in the previous paper\cite{sakatake2022}.
For those who are familiar with it, readers can directly go to the next section.
By taking up the quadratic terms in the shifted Lagrangian around the constant field $\varphi$
for the scalar field $\phi$ in Eq.(\ref{2.1}), one needs to evaluate 
\begin{align}
V_{\rm eff}
&=(-1)^{f}{\cal N}~\half\left( \prod_{i=0}^{p}\frac{1}{L_i}\sum_{n_i=-\infty}^{\infty}\right)
\int\frac{d^{D-(p+1)}p_E}{(2\pi)^{D-(p+1)}}
   \notag\\
&\hspace{5mm}\times
\log\Bigl[p_E^2 +\left(\frac{2\pi}{L_0}\right)^2(n_0 +\eta_0)^2
+\sum_{i=1}^p\left(\frac{2\pi}{L_i}\right)^2(n_i +\eta_i)^2
+M^2(\varphi)\Bigr]
 \label{2.4}
\end{align}
in order to obtain the effective potential on $S_{\tau}^1\times R^{D-(p+1)}\times \prod_{i=1}^{p}S^1_i$
in one-loop approximation. Here, the $M(\varphi)$ is the field-dependent mass 
of the scalar (fermion) field, 
\begin{align}
M^2(\varphi)=m_s^2+\frac{\lambda}{2}\varphi^2 \quad \Bigl(M(\varphi)=m_F+g\varphi\Bigr).
 \label{2.5}
\end{align}
Hereafter, we denote $M(\varphi)$ by $M$ for simplicity. The $p_E$ denotes the $D-(p+1)$-dimensional 
Euclidean momentum. The $f$ is the fermion number that is $0~(1)$ for the boson (fermion) and 
the ${\cal N}$ is the on-shell degrees of freedom. The $n_{0}$ denotes the Matsubara mode arising 
from the $S_{\tau}^1$, and the Kaluza-Klein mode $n_i~(i=1,\cdots, p)$ comes from each $S_i^1~(i=1,\cdots p)$. The 
parameter $\eta_i\, (i=0,1,\cdots, p)$ is given in Eqs.(\ref{2.2}) and (\ref{2.3}).

Let us make use of the zeta-function regularization in order to evaluate Eq.(\ref{2.4}). By defining 
\begin{align}
I(s)&\equiv \left( \prod_{i=0}^{p}\frac{1}{L_i}\sum_{n_i=-\infty}^{\infty}\right)
\int\frac{d^{D-(p+1)}p_E}{(2\pi)^{D-(p+1)}}\notag\\
&\hspace{5mm}
\times 
\Bigl[p_E^2 +\left(\frac{2\pi}{L_0}\right)^2(n_0 +\eta_0)^2
+\sum_{i=1}^p\left(\frac{2\pi}{L_i}\right)^2(n_i +\eta_i)^2
+M^2\Bigr]^{-s},
 \label{2.6}
\end{align}
then, the $V_{\rm eff}$ is written as
\begin{align}
V_{\rm eff}=(-1)^f{\cal N}\half\left(-\frac{d}{ds}I(s)\right)\Big|_{s\rightarrow 0}.
 \label{2.7}
\end{align}
Performing the $p_E$ integration with the formula
\begin{align}
A^{-s}=\frac{1}{\Gamma(s)}\int_0^{\infty}dt~t^{s-1}\e^{-At},
 \label{2.8}
\end{align} 
%
%we perform the $p_E$ integration, 
%
%
and employing the Poisson summation
\begin{align}
\sum_{n_j=-\infty}^{\infty}\e^{-(\frac{2\pi}{L_j})^2(n_j+\eta_j)^2t}=
\sum_{m_j=-\infty}^{\infty}\frac{L_j}{2\pi}\left(\frac{\pi}{t}\right)^{\half}
\e^{-\frac{(m_jL_j)^2}{4t}+2\pi im_j\eta_j},
 \label{2.9}
\end{align}
we obtain
\begin{align}
V_{\rm eff}&=(-1)^{f+1}{\cal N}\frac{\pi^{\frac{D}{2}}}{2(2\pi)^D}
\sum_{m_0=-\infty}^{\infty}\cdots \sum_{m_p=-\infty}^{\infty}
\notag \\
&\hspace{5mm}\times
  \int_0^{\infty}dt~t^{-\frac{D}{2}-1}\e^{-\frac{1}{4t}[(m_0L_0)^2+\cdots+(m_pL_p)^2] -M^2t +2\pi i(m_0\eta_0+\cdots +m_p\eta_p)}.
   \label{2.10}
\end{align}
Hereafter, we call $m_j~(j=0, 1,\cdots, p)$ in Eq.(\ref{2.9}) or Eq.(\ref{2.10}) the winding modes, while $n_0$ and $n_i~(i=1,2,\cdots, p)$
in Eq.(\ref{2.9}) or Eq.(\ref{2.4}) the Matsubara and the Kaluza-Klein modes, respectively.

It is useful to separate each summation with respect to $m_j$ in Eq.(\ref{2.10}) into the zero 
mode $(m_j=0)$ and the nonzero ones $(m_j\neq 0)$ and to express Eq.(\ref{2.10}) into the  form
%%%%%%%%%%%%%
\begin{align}
V_{\text{eff}} = \sum_{n=0}^{p+1} F^{(n)D}(M)=
\sum_{n=0}^{p+1}~\sum_{0\leq i_1<i_2<\cdots < i_n\leq p}F^{(n)D}_{L_{i_1},L_{i_2},\cdots, L_{i_n}}(M),
 \label{2.11}
\end{align}
where
\begin{align}
F^{(n)D}_{L_{i_1},L_{i_2},\cdots, L_{i_n}}(M)
 &= (-1)^{f+1}{\cal N}\frac{\pi^{\frac{D}{2}}}{2(2\pi)^D}
     \sum_{m_{i_1}=-\infty}^{\infty}{}^{\hspace{-3mm}\prime}\cdots 
     \sum_{m_{i_n}=-\infty}^{\infty}{}^{\hspace{-3mm}\prime}
     \notag \\
 &\hspace{5mm}\times
    \int_0^{\infty}dt~t^{-\frac{D}{2}-1}\e^{-\frac{1}{4t}[(m_{i_1}L_{i_1})^2+
     \cdots+(m_{i_n}L_{i_n})^2]-M^2t +2\pi i(m_{i_1}\eta_{i_1}+\cdots +m_{i_n}\eta_{i_n})}.
      \label{2.12}
\end{align}
%%%%%%%%%%%%%
The prime of the summation in $\sum_{m_j=-\infty}^{\,\prime\, \infty}$ means that the zero mode ($m_j=0$) is removed.

The $F^{(0)}$ in Eq.(\ref{2.11}) corresponds to the contribution from
all the zero modes $m_{0} = m_{1} = \cdots = m_{p} = 0$ in Eq.(\ref{2.10}) and is found to be
%%%%%%%%%%%%%
\begin{align}
F^{(0)D}(M)
 = (-1)^{f+1}{\cal N}\frac{\pi^{\frac{D}{2}}}{2(2\pi)^D}
    \int_0^{\infty}dt~t^{-\frac{D}{2}-1}\e^{-M^2t}
 = (-1)^{f+1}{\cal N}\frac{\pi^{\frac{D}{2}}}{2(2\pi)^D}
   \Gamma(-\tfrac{D}{2})(M^2)^{\frac{D}{2}}.
     \label{2.13}
\end{align}
%%%%%%%%%%%%%
It must be understood that $F^{(0)D}(M)$ is regularized by the dimensional 
regularization for $D=$ even. On the other hand,  for $D=$ odd, it yields\cite{sakatake2022}
\begin{align}
%
%F^{(0)D}(M)=\frac{-(-1)^{\frac{D+1}{2}}}{2^{\frac{D+1}{2}}\pi^{\frac{D-1}{2}}D!!}M^D.
%
F^{(0)D}(M)=(-1)^{f+1}{\cal N}\frac{(-1)^{\frac{D+1}{2}}}{2^{\frac{D+1}{2}}\pi^{\frac{D-1}{2}}D!!}M^D.
  \label{2.14}
\end{align}
By using the formula 
\begin{align}
\int_0^{\infty}dt~t^{-\nu-1}\e^{-At-\frac{B}{t}}=2\left(\frac{A}{B}\right)^{\frac{\nu}{2}}K_{\nu}(2\sqrt{AB}),
  \label{2.15}
\end{align}
where the $K_{\nu}(z)$ is the modified Bessel function of the second kind, Eq.(\ref{2.12}) for $n\geq 1$ becomes
\begin{align}
F_{L_{i_1},L_{i_2},\cdots, L_{i_{n}}}^{(n)D}(M)
 &= (-1)^{f+1}{\cal N}\frac{2^n}{(2\pi)^{\frac{D}{2}}}
    \sum_{m_{i_{1}}=1}^{\infty} \cdots \sum_{m_{i_{n}}=1}^{\infty}
    \left(\frac{M^2}{(m_{i_1}L_{i_1})^2+\cdots +(m_{i_n}L_{i_n})^2}\right)^{\frac{D}{4}}
    \notag\\
&\times 
    K_{\frac{D}{2}}\left(\sqrt{{M^2}\{(m_{i_1}L_{i_1})^2+\cdots +(m_{i_n}L_{i_n})^2\}}\right)
    \cos(2\pi m_{i_1}\eta_{i_1})\cdots \cos(2\pi m_{i_n}\eta_{i_n}).
      \label{2.16}
\end{align}
%%%%%%%%%%%

The effective potential is given by the modified Bessel function of the second kind accompanied with 
the multiple mode summation with respect to the winding mode $m_{i_j} (j=1,\cdots, n)$ associated with 
the $S_{i_j}^1$. The winding mode is derived from the Matsubara/Kaluza-Klein mode $n_{i_j}$ through the 
Poisson summation (\ref{2.9}). We will use the inverse process in the next 
section, which is employed to prove a mode recombination formula. The results in this section have already been  
obtained in the previous paper (see Eq.(2.10) of Ref.\cite{sakatake2022}) and they are the starting point 
for the discussion in the present paper. 
%
%
%
%
%%%%%%%%%%%%%%%%%%%%%%%%%%%%%
\section{Mode Recombination Formula}
%%%%%%%%%%%%%%%%%%%%%%%%%%%%%
%
%
%
%
In this section, we present and prove the formula called a mode recombination formula. The formula plays a crucial role to 
obtain a new form of the effective potential, which is different from Eq.(\ref{2.11}) and is convenient for discussing the nonanalytic 
terms. These terms can be obtained from the effective potential by use of the mode recombination formula
in easier and more transparent way, as we will see in the next section.

Let us recall Eq.(\ref{2.11}) and first write it as
\begin{align}
V_{\rm eff}=F^{(0)D}(M)+\sum_{n=1}^{p+1}~\sum_{0\leq i_1<i_2<\cdots < i_n\leq p}F^{(n)D}_{L_{i_1},L_{i_2},\cdots, L_{i_n}}(M),
\label{3.1}
\end{align}
where we have separated the $n=0$ term from the summation in Eq.(\ref{2.11}) for later convenience.
We focus on the scale $L_0$ in $L_{i_j} (i_j=0,\cdots,p)$ and separate the term associated with 
$L_0$ from the terms without $L_0$ on the right-hand side of Eq.(\ref{3.1}) 
as\footnote{Let us comment that the effective potential obtained by 
Dolan-Jackiw, Eq.(3.13c) in their paper\cite{dj}, is reduced to $F_{L_0}^{(1)D}(M)$ by applying the formula\cite{bromwich}
\begin{align}
K_{\half}(z)=\sqrt{\frac{\pi}{2z}}\e^{-z}\quad{\rm and}\quad 
\int_0^{\infty}dx~K_{\nu}(\alpha\sqrt{x^2+z^2})\frac{x^{2\mu +1}}{(x^2+z^2)^{\frac{\nu}{2}}}
=\frac{2^{\mu}\Gamma(\mu+1)}{\alpha^{\mu+1}z^{\nu-\mu-1}}K_{\nu-\mu-1}(\alpha z)
\notag
\end{align}
to the series expansion of the logarithm in Eq.(3.13c). The mode in the series expansion plays the same role with 
the winding mode $m_0$ in $F_{L_0}^{(1)D}(M)$.}
%
%\newpage\noindent
%
%
\begin{align}
&\sum_{n=1}^{p+1}~\sum_{0\leq i_1<i_2<\cdots < i_n\leq p}F^{(n)D}_{L_{i_1},L_{i_2},\cdots, L_{i_n}}(M)\notag\\
&\hspace{5mm}
=F_{L_0}^{(1)D}(M)+
\sum_{n=1}^{p}~\sum_{1\leq i_1<i_2<\cdots < i_n\leq p}
\Bigl(F_{L_{i_1},\cdots, L_{i_n}}^{(n)D}(M)+F_{L_0, L_{i_1},\cdots, L_{i_n}}^{(n+1)D}(M)\Bigr).
  \label{3.2}
\end{align}
Let us show that the second and third terms on the right-hand side of Eq.(\ref{3.2}) can be combined into a single expression
\begin{align}
F_{L_{i_1},\cdots, L_{i_n}}^{(n)D}(M)+F_{L_0, L_{i_1},\cdots, L_{i_n}}^{(n+1)D}(M)=
\frac{1}{L_0}\sum_{n_0=-\infty}^{\infty}F_{L_{i_1},\cdots, L_{i_n}}^{(n)D-1}(M_{(0)}),
  \label{3.3}
\end{align}
where 
\begin{align}
M_{(0)}^2\equiv M^2+\left(\frac{2\pi}{L_0}\right)^2(n_0+\eta_0)^2.
  \label{3.4}
\end{align}
Here, the $n_0$ is the original Matsubara mode appeared in Eq.(\ref{2.4}).

Let us prove Eq.(\ref{3.3}). By using Eq.({\ref{2.16}), the two terms on the left-hand side of Eq.(\ref{3.3}) are combined 
into a single expression, including 
the $m_0=0$ mode\footnote{Let us note that this is allowed 
because Eq.(\ref{2.16}) is regularized with respect to $m_{i_j}=0~(j=1,\cdots, n)$.}, as
\begin{align}
&F_{L_{i_1},\cdots, L_{i_n}}^{(n)D}(M)+F_{L_0, L_{i_1},\cdots, L_{i_n}}^{(n+1)D}(M)\notag\\
&\hspace{5mm}=
(-1)^{f+1}{\cal N}\frac{2^n}{(2\pi)^{\frac{D}{2}}}\sum_{m_0=-\infty}^{\infty}
\sum_{m_{i_1}=1}^{\infty}\cdots \sum_{m_{i_n}=1}^{\infty}
\left(\frac{M^2}{(m_0L_0)^2+(m_{i_1}L_{i_1})^2+\cdots+(m_{i_n}L_{i_n})^2}\right)^{\frac{D}{4}}
\notag\\
&\hspace{5mm}\times K_{\frac{D}{2}}\left(\sqrt{M^2\left\{(m_0L_0)^2+(m_{i_1}L_{i_1})^2+\cdots+(m_{i_n}L_{i_n})^2\right\}}\right)
\e^{2\pi i m_0\eta_0}
\cos(2\pi m_{i_1}\eta_{i_1})\cdots \cos(2\pi m_{i_n}\eta_{i_n}).
  \label{3.5}
\end{align}
With the help of the formula (\ref{2.15}), the modified Bessel function of the 
second kind in Eq.(\ref{3.5}) is written in the integral form, and we inversely use the Poisson 
summation (\ref{2.9}) for the $m_0$ mode, which is then converted into the Matsubara mode $n_0$. Thus, we obtain
\begin{align}
&F_{L_{i_1},\cdots, L_{i_n}}^{(n)D}(M)+F_{L_0, L_{i_1},\cdots, L_{i_n}}^{(n+1)D}(M)\notag\\
&\hspace{5mm}=
\frac{1}{L_0}(-1)^{f+1}{\cal N}\frac{2^n}{2(4\pi)^{\frac{D-1}{2}}}\sum_{n_0=-\infty}^{\infty}\sum_{m_{i_1}=1}^{\infty}\cdots 
\sum_{m_{i_n}=1}^{\infty}\int_0^{\infty}dt~t^{-\frac{D-1}{2}-1}\notag\\
&\hspace{5mm}\times 
\e^{-[M^2+(\frac{2\pi}{L_0})^2(n_0+\eta_0)^2]t -\frac{1}{4t}[(m_{i_1}L_{i_1})^2+\cdots +(m_{i_n}L_{i_n})^2]}
\cos(2\pi m_{i_1}\eta_{i_1})\cdots \cos(2\pi m_{i_n}\eta_{i_n})\notag\\
&\hspace{5mm}=\frac{1}{L_0}
(-1)^{f+1}{\cal N}\frac{2^n}{(2\pi)^{\frac{D-1}{2}}}\sum_{n_0=-\infty}^{\infty}
\sum_{m_{i_1}=1}^{\infty}\cdots \sum_{m_{i_n}=1}^{\infty}\left(\frac{M_{(0)}^2}{(m_{i_1}L_{i_1})^2+\cdots+(m_{i_n}L_{i_n})^2}\right)^{\frac{D-1}{4}}
\notag\\
&\hspace{5mm}\times K_{\frac{D-1}{2}}\left(\sqrt{M_{(0)}^2\left\{(m_{i_1}L_{i_1})^2+\cdots+(m_{i_n}L_{i_n})^2\right\}}\right)
\cos(2\pi m_{i_1}\eta_{i_1})\cdots \cos(2\pi m_{i_n}\eta_{i_n}),
 \label{3.6}
\end{align}
where we have used Eq.(\ref{2.15}) again to rewrite the integral from into the Bessel function 
in the last equality. Eq.(\ref{3.6}) is nothing but the right-hand side of Eq.(\ref{3.3}) and we have 
proved Eq.(\ref{3.3}).

We shall call Eq.(\ref{3.3}) the mode recombination formula\footnote{The idea of the mode recombination 
has been developed in a different context by the authors\cite{sakatake}.} in the present paper.
Let us note that the Matsubara mass squared, the second term in Eq.(\ref{3.4}), is recovered by the 
inverse use of the Poisson summation (\ref{2.9}). From Eqs.(\ref{3.2}) and (\ref{3.3}), we obtain an important relation
\begin{align}
&\sum_{n=1}^{p+1}~\sum_{0\leq i_1<i_2<\cdots < i_n\leq p}F^{(n)D}_{L_{i_1},L_{i_2},\cdots, L_{i_n}}(M)\notag\\
&\hspace{5mm}
=
F_{L_0}^{(1)D}(M)+\sum_{n=1}^{p}~\sum_{1\leq i_1<i_2<\cdots < i_n\leq p}
\frac{1}{L_0}\sum_{n_0=-\infty}^{\infty}F_{L_{i_1},\cdots, L_{i_n}}^{(n)D-1}(M_{(0)}).
 \label{3.7}
\end{align}

Some comments are in order. It is important to note that 
the first term on the right-hand side of Eq.(\ref{3.7}), which is the contribution of the 
single mode summation with respect to the winding mode $m_0$ associated with 
the $S_{\tau}^1$ having the focused scale $L_0$, is separated from the contributions of the multiple mode summations in the process 
of deriving the mode recombination formula. The Matsubara mass squared, the second term in Eq.(\ref{3.4}), arises after using inversely 
the Poisson summation. As is clear from the discussions given above, the formula (\ref{3.7}) holds irrespective of whether the field 
is the scalar or fermion and also of the boundary condition for the spatial $S_i^1~(i=1,\cdots, p)$ direction. The spacetime dimensions 
in $F_{L_{i_1},\cdots, L_{i_n}}^{(n)D-1}(M_{(0)})$ is effectively reduced to $D-1$. This is interpreted as that the particle with 
mass squared $M^2$ having the Matsubara mass squared $(\tfrac{2\pi}{L_0})^2(n_0+\eta_0)^2$ to be considered on 
the $D-1$ dimensional spacetime, $R^{D-(p+1)}\times \prod_{i=1}^pS^1_i$. One can choose another 
scale, say $L_j~(j\in \{1,\cdots, p\})$ instead of $L_0$, then,  the $M_{(0)}^2$ is replaced by $M_{(j)}^2$ with the Kaluza-Klein 
mass squared $(\tfrac{2\pi}{L_j})^2(n_j+\eta_j)^2$.

From Eq.(\ref{3.7}), the effective potential (\ref{3.1}) is written by
\begin{align}
V_{\rm eff}=F^{(0)D}(M)+F_{L_0}^{(1)D}(M)
+\sum_{n=1}^{p}~\sum_{1\leq i_1<i_2<\cdots < i_n\leq p}\frac{1}{L_0}\sum_{n_0=-\infty}^{\infty}F_{L_{i_1},\cdots, L_{i_n}}^{(n)D-1}(M_{(0)}).
 \label{3.8}
\end{align}
We repeat the same discussion for the third term in Eq.(\ref{3.8}) by focusing next on the scale $L_1$. 
One immediately sees that the term, aside from the scale $L_0$, can be written as 
\begin{align}
&\sum_{n=1}^{p}~\sum_{1\leq i_1<i_2<\cdots < i_n\leq p}~\sum_{n_0=-\infty}^{\infty}F_{L_{i_1},\cdots, L_{i_n}}^{(n)D-1}(M_{(0)})
\notag\\
&\hspace{5mm}
=\sum_{n_0=-\infty}^{\infty}F_{L_1}^{(1)D-1}(M_{(0)})+
\sum_{n=1}^{p-1}~\sum_{2\leq i_1<i_2<\cdots < i_n\leq p}~\sum_{n_0=-\infty}^{\infty}
\left(
F_{L_{i_1},\cdots,L_{i_n}}^{(n)D-1}(M_{(0)})+
F_{L_1,L_{i_1},\cdots,L_{i_n}}^{(n+1)D-1}(M_{(0)})
\right)
\notag\\
&\hspace{5mm}
=\sum_{n_0=-\infty}^{\infty}F_{L_1}^{(1)D-1}(M_{(0)})+
\sum_{n=1}^{p-1}~\sum_{2\leq i_1<i_2<\cdots < i_n\leq p}~\sum_{n_0=-\infty}^{\infty}~\frac{1}{L_1}\sum_{n_1=-\infty}^{\infty}
F_{L_{i_1},\cdots, L_{i_n}}^{(n)D-2}(M_{(0,1)}),
 \label{3.9}
\end{align}
where we have defined 
\begin{align}
M_{(0,1)}^2=M^2+\sum_{i=0}^1\left(\frac{2\pi}{L_i}\right)^2(n_i+\eta_i)^2
 \label{3.10}
\end{align}
and have used Eq.(\ref{3.3}) with $L_0, D-1, M$ and $M_{(0)}$ being replaced by $L_1, D-2, M_{(0)}$ and $M_{(0,1)}$, respectively 
in the last equality. We observe that the first term on the right-hand side of Eq.(\ref{3.9}), the contribution of 
the single mode summation with respect to the winding mode $m_1$ associated with the $S_1^1$ having the focused 
scale $L_1$, is separated from the contributions of the multiple mode summations. The Kaluza-Klein 
mass squared $(\tfrac{2\pi}{L_1})^2(n_1+\eta_1)^2$ turns out to be added to $M_{(0)}^2$ 
through the formula (\ref{3.3}) in Eq.(\ref{3.10}).
%
%part of the multiple mode summations in Eq.(3.9).
%
%

We repeat the same discussion given above by focusing on the scales $L_2, L_3,\cdots, L_{p-1}$ sequentially for the
multiple mode summation obtained at each step such as the second term on the right-hand side of Eq.(\ref{3.7}) (or Eq.(\ref{3.9})). 
We finally find that the effective potential is recast into
\begin{align}
V_{\rm eff}&=F^{(0)D}(M)+F_{L_0}^{(1)D}(M)+\frac{1}{L_0}\sum_{n_0=-\infty}^{\infty}F_{L_1}^{(1)D-1}(M_{(0)})+
\frac{1}{L_0L_1}\sum_{n_0=-\infty}^{\infty}\sum_{n_1=-\infty}^{\infty}F_{L_2}^{(1)D-2}(M_{(0,1)})\notag\\
&\hspace{5mm}+\cdots +
\frac{1}{L_0\cdots L_{p-1}}\sum_{n_0=-\infty}^{\infty}\cdots 
\sum_{n_{p-1}=-\infty}^{\infty}F_{L_p}^{(1)D-p}(M_{(0,1,\cdots, p-1)})\notag\\
&=F^{(0)D}(M)+\sum_{k=0}^p\left(\frac{1}{\prod_{i=0}^{k-1}L_i}\right)\left(\prod_{j=0}^{k-1}\sum_{n_j=-\infty}^{\infty}\right)
F_{L_k}^{(1)D-k}(M_{(0,1,\cdots,k-1)}),
 \label{3.11}
\end{align}
where we have defined 
\begin{align}
M_{(0,1,\cdots,k-1)}^2\equiv M^2+\sum_{i=0}^{k-1}\left(\frac{2\pi}{L_i}\right)^2(n_i+\eta_i)^2.
 \label{3.12}
\end{align}
The second term $F_{L_0}^{(1)D}(M)$ in the first line of Eq.(\ref{3.11}) is incorporated into the summation
with respect to $k$ as the $k=0$ term in Eq.(\ref{3.11}), where it is understood that we set 
\begin{align}
\prod_{i=0}^{k-1}L_i\Bigg|_{k=0}=\prod_{j=0}^{k-1}\sum_{n_j=-\infty}^{\infty}\Bigg|_{k=0}=1,\quad 
M_{(0,1,\cdots,k-1)}^2\Big|_{k=0}=M^2.
 \label{3.13}
\end{align}

Eq.(\ref{3.11}) is the effective potential rewritten by using successively the mode recombination formula (\ref{3.3}) and 
gives the starting point to discuss the nonanalytic terms in the present paper. Let us note that Eq.(\ref{3.11}) holds for 
any boundary condition for the spatial $S_i^1$ direction, reflecting the fact that the mode recombination formula 
holds irrespective of the boundary condition for any spatial direction.

 It may be instructive to mention that the effective potential (\ref{3.1}) is invariant under the exchange of the 
scales $L_i$ and $L_j~(i\neq j)$. The invariance, however, does not become manifest in the new form of the 
effective potential (\ref{3.11}). Instead of losing the manifest invariance, the effective potential has a remarkable feature that except for the first 
term in Eq.(\ref{3.11}), all the contributions to the effective potential are given by the single mode summation with respect to the 
winding mode $m_k (k=0,\cdots, p)$ associated with the $S_k^1$, even though 
the $M^2$ has the Kaluza-Klein (Matsubara) mass squared, the second term in Eq.(\ref{3.12}).
One can further simplify Eq.(\ref{3.11}) by taking account of the discussion on the zero modes in Eq.(\ref{3.12}), as we will see in the next section.

We shall study the nonanalytic terms for the case of the fermion field as well as the 
scalar one in the present paper. It may be appropriate here to present 
the convenient form of the effective potential for the discussion in the case of the fermion. 
One can show that the first and the second terms in Eq.(\ref{3.8}) are combined into a single expression
 \begin{align}
 F^{(0)D}(M)+F_{L_0}^{(1)D}(M)=\frac{1}{L_0}\sum_{n_0=-\infty}^{\infty}F^{(0)D-1}(M_{(0)}).
 \label{3.14}
 \end{align}
Then, the effective potential (\ref{3.8}) is written as
 \begin{align}
 V_{\rm eff}&=\frac{1}{L_0}\sum_{n_0=-\infty}^{\infty}F^{(0)D-1}(M_{(0)})
+\sum_{n=1}^{p}~\sum_{1\leq i_1<i_2<\cdots < i_n\leq p}\frac{1}{L_0}\sum_{n_0=-\infty}^{\infty}F_{L_{i_1},\cdots, L_{i_n}}^{(n)D-1}(M_{(0)})
\label{3.15}
 \end{align}
or
 \begin{align}
V_{\rm eff}=\sum_{n=0}^{p}~\sum_{1\leq i_1<i_2<\cdots < i_n\leq p}\frac{1}{L_0}\sum_{n_0=-\infty}^{\infty}F_{L_{i_1},\cdots, L_{i_n}}^{(n)D-1}(M_{(0)}),
\label{3.16}
 \end{align}
where the first term in Eq.(\ref{3.15}) is incorporated into the summation with respect to $n$ as the $n=0$ term in Eq.(\ref{3.16}).

Let us prove Eq.(\ref{3.14}). By setting $m_{i_1}=\cdots=m_{i_n}=0, n=0$ and dropping the 
summation $\sum_{m_j}~(j=1,\cdots, n)$ in the first equality of Eq.(\ref{3.6}), the left-hand side of Eq.(\ref{3.14}) is given by
\begin{align}
%
%F^{(0)D}(M)+F_{L_0}^{(1)}(M)
%
%\frac{1}{L_0}\sum_{n_0=-\infty}^{\infty}F^{(0)D-1}(M_{(0)})
%
F^{(0)D}(M)+F_{L_0}^{(1)D}(M)
&=\frac{1}{L_0}(-1)^{f+1}{\cal N}\frac{1}{2(4\pi)^{\frac{D-1}{2}}}
\sum_{n_0=-\infty}^{\infty}
\int_0^{\infty}dt~t^{-\frac{D-1}{2}-1}\e^{-[M^2+(\frac{2\pi}{L_0})^2(n_0+\eta_0)^2]t }\notag\\
&=\frac{1}{L_0}(-1)^{f+1}{\cal N}\frac{\pi^{\frac{D-1}{2}}}{2(2\pi)^{D-1}}\Gamma(-\tfrac{D-1}{2})
\sum_{n_0=-\infty}^{\infty}\left(M_{(0)}^2\right)^{\frac{D-1}{2}},
\label{3.17}
\end{align}
where we have used Eq.(\ref{2.8}) in the last equality. From Eq.(\ref{2.13}), this is nothing but the right-hand side of Eq.(\ref{3.14}) and 
we have proved Eq.(\ref{3.14}).

We will use Eqs.(\ref{3.15}) and (\ref{3.16}) in section $5$, where we discuss the nonanalytic terms in the effective potential
for the case of the fermion. It can be said that Eq.(\ref{3.14}) is also the mode recombination formula 
and is regarded as the $n=0$ case in Eq.(\ref{3.3}). Let us note that we can always write the effective potential 
as Eq.(\ref{3.16}) for any boundary condition of the spatial $S_i^1$ direction.

For later convenience, we introduce the integral representation for the modified 
Bessel function of the second kind on the complex plane\cite{bromwich, davis}
\begin{align}
K_{\nu}(x)=\frac{1}{4\pi i}\int_{c-i\infty}^{c+i\infty}dt~\Gamma(t)\Gamma(t-\nu)\left(\frac{x}{2}\right)^{-2t+\nu}.
\label{3.18}
\end{align}
The constant $c$ should be understood to be a point located on the real axis which is greater than all the poles of the gamma 
functions in the integrand. By using the formula (\ref{3.18}), one obtains the integral form of the right-hand side of Eq.(\ref{3.3}) with $n\geq 1$ as
\begin{align}
&\frac{1}{L_0}\sum_{n_0=-\infty}^{\infty}F_{L_{i_1},\cdots, L_{i_n}}^{(n)D-1}(M_{(0)})\notag\\
&\hspace{5mm}
=\frac{1}{L_0}(-1)^{f+1}{\cal N}\frac{2^n}{(2\pi)^{\frac{D-1}{2}}}\sum_{n_0=-\infty}^{\infty}
\left(\frac{M_{(0)}^2}{2}\right)^{\frac{D-1}{2}}
\frac{1}{4\pi i}
\int_{c-i\infty}^{c+i\infty}dt~\Gamma(t-\tfrac{D-1}{2})\left(\frac{M_{(0)}}{2}\right)^{-2t}\notag\\
&\hspace{10mm}\times
S^{(n)}(t; L_{i_1},\cdots, L_{i_n}),
\label{3.19}
\end{align}
where we have defined 
\begin{align}
S^{(n)}(t; L_{i_1},\cdots, L_{i_n})&\equiv \Gamma(t)\sum_{m_{i_1}=1}^{\infty}\cdots\sum_{m_{i_n}=1}^{\infty}
\left\{
(m_{i_1}L_{i_1})^2+\cdots +(m_{i_n}L_{i_n})^2
\right\}^{-t}\notag\\
&\hspace{5mm}\times \cos(2\pi m_{i_1}\eta_{i_1})\cdots \cos(2\pi m_{i_n}\eta_{i_n}),
\label{3.20}
\end{align} 
and $M_{(0)}$ is given by Eq.(\ref{3.4}). Let us note that the $n=0$ case in Eq.(\ref{3.3}) is given by Eq.(\ref{3.17}).

We have succeeded in representing the effective potential in the various forms
like Eqs.(\ref{3.8}), (\ref{3.11}), (\ref{3.15}) and (\ref{3.16}). It may be helpful to comment on which forms 
are useful for the analysis of the scalar or fermion field. The expressions (\ref{3.8}) and (\ref{3.11}) 
of the effective potential will be used for the scalar with the periodic boundary 
condition. On the other hand, the expressions (\ref{3.15}) and (\ref{3.16}) turn out to be 
useful for the fermion and the scalar with the antiperiodic boundary condition.

%%%%%%%%%%%%%%%%%%%%%%%%%%%%%%%%%%%%%%
\section{Nonanalytic terms for scalar field with periodic boundary condition} 
%%%%%%%%%%%%%%%%%%%%%%%%%%%%%%%%%%%%%%
%
%
%
%
In this section, we derive the nonanalytic terms of the effective potential for a real scalar $(f=0, {\cal N}=1, \eta_0=0)$
with the periodic boundary condition in all the spatial directions $(\eta_1=\cdots=\eta_p=0)$, although the
results have already been obtained in the previous paper\cite{sakatake2022}. The purpose of this section 
is to show that our formulation presented in this paper is easier and more transparent to 
obtain the nonanalytic terms than that in the previous paper. To 
this end, we examine the right-hand side of Eq.(\ref{3.3}). From Eq.(\ref{3.19}), we have 
\begin{align}
&\frac{1}{L_0}\sum_{n_0=-\infty}^{\infty}F_{L_{i_1},\cdots, L_{i_n}}^{(n)D-1}(M_{(0)})\notag\\
&\hspace{5mm}
%
%
%=\frac{-1}{L_0}\sum_{n_0=-\infty}^{\infty}\frac{2^n}{(2\pi)^{\frac{D-1}{2}}}
%\left(\frac{M_{(0)}^2}{2}\right)^{\frac{D-1}{2}}
%\frac{1}{4\pi i}
%\int_{c-i\infty}^{c+i\infty}dt~\Gamma(t-\tfrac{D-1}{2})\left(\frac{M_{(0)}}{2}\right)^{-2t}S^{(n)}(t; L_{i_1},\cdots, L_{i_n})
%\notag\\
%&\hspace{5mm}
%
=-\frac{2^n}{(2\pi)^{\frac{D-1}{2}}L_0}\left(\frac{M^2}{2}\right)^{\frac{D-1}{2}}\frac{1}{4\pi i}
\int_{c-i\infty}^{c+i\infty}dt~\Gamma(t-\tfrac{D-1}{2})\left(\frac{M}{2}\right)^{-2t}{\tilde S}^{(n)}(t; L_{i_1},\cdots, L_{i_n})
\notag\\
&\hspace{5mm}
-\frac{2^{n+1}\pi^{\frac{D-1}{2}}}{L_0^{D}}\frac{1}{4\pi i}\int_{c-i\infty}^{c+i\infty}dt~
\Gamma(t-\tfrac{D-1}{2})\sum_{n_0=1}^{\infty}
\left\{
n_0^2+\left(\frac{ML_0}{2\pi}\right)^2\right\}^{\tfrac{D-1}{2}-t}\left(\frac{\pi}{L_0}\right)^{-2t}
{\tilde S}^{(n)}(t; L_{i_1},\cdots, L_{i_n}),
\label{4.1}
\end{align}
where we have defined 
\begin{align}
{\tilde S}^{(n)}(t; L_{i_1},\cdots, L_{i_n})&\equiv \Gamma(t)\sum_{m_{i_1}=1}^{\infty}\cdots\sum_{m_{i_n}=1}^{\infty}
\left\{
(m_{i_1}L_{i_1})^2+\cdots +(m_{i_n}L_{i_n})^2
\right\}^{-t}
\label{4.2}
\end{align} 
and have separated the zero mode $(n_0=0)$ from the nonzero modes $(n_0\neq 0)$.
%
%, which is given by the first term and the second term in (4.2), respectively.
%

We show that the second term in Eq.(\ref{4.1}) never has nonanalytic 
terms, in other words, all the terms are given by positive integer powers of 
the mass squared $M^2$. The second term in Eq.(4.1) is written, after changing the variable $\bar t =t-\tfrac{D-1}{2}$, as
\begin{align}
-\frac{2^{n+1}}{\pi^{\frac{D-1}{2}}L_0}\frac{1}{4\pi i}\int_{\bar c-i\infty}^{\bar c+i\infty}dt~
\Gamma(t)\sum_{n_0=1}^{\infty}
\left\{
n_0^2+\left(\frac{ML_0}{2\pi}\right)^2\right\}^{-t}
\left(\frac{\pi}{L_0}\right)^{-2t}{\tilde S}^{(n)}(t+\tfrac{D-1}{2}; L_{i_1},\cdots, L_{i_n}),
\label{4.3}
\end{align}
where we have again denoted $\bar t$ by $t$. Let us recall the following formula
used in the previous paper\cite{sakatake2022,elizalde}
\begin{align}
&\Gamma(t)\sum_{n_0=1}^{\infty}\{(n_0L_0)^2+z^2\}^{-t}\nonumber\\
&=
-\half  \frac{\Gamma(t)}{z^{2t}} 
+ \frac{\sqrt{\pi}}{2L_0}\frac{\Gamma(t-\half)}{z^{2(t-\half)}}
+\frac{2\pi^t}{L_0^{t+\half}}\frac{1}{z^{t-\half}}
\sum_{m=1}^{\infty}m^{t-\half}~K_{t-\half}\left(\frac{2\pi m}{L_0}z\right)
\nonumber\\
&=
-\half  \frac{\Gamma(t)}{z^{2t}} 
+ \frac{\sqrt{\pi}}{2L_0}\frac{\Gamma(t-\half)}{z^{2(t-\half)}}\nonumber\\
&\hspace{5mm}
+
\frac{1}{\sqrt{\pi}}\left(\frac{\pi}{L_0}\right)^{2t}
\frac{1}{2\pi i}\int_{c_{1} - i\infty}^{c_{1} +i\infty}d{t_{1}}~\Gamma(t_{1} -t +\tfrac{1}{2})
\zeta(2t_{1}-2t +1)\Gamma(t_{1}) \left(z\frac{\pi}{L_0}\right)^{-2t_{1}},
\label{4.4}
\end{align}
where we have used Eq.(\ref{3.18}) in the last equality.

We apply the formula (\ref{4.4}) with $L_0=1$ and $z=\tfrac{ML_0}{2\pi}$ to Eq.(\ref{4.3}), which becomes  
\begin{align}
&-\frac{2^{n+1}}{\pi^{\frac{D-1}{2}}L_0}\frac{1}{4\pi i}\int_{\bar c-i\infty}^{\bar c+i\infty}dt~
\Biggl\{
-\half \Gamma(t)\left(\frac{ML_0}{2\pi}\right)^{-2t}+\frac{\sqrt{\pi}}{2}
\Gamma(t-\tfrac{1}{2})\left(\frac{ML_0}{2\pi}\right)^{-2(t-\frac{1}{2})}
\notag\\
&\hspace{5mm}+
\frac{\pi^{2t}}{\sqrt{\pi}}
\frac{1}{2\pi i}\int_{c_1-i\infty}^{c_1+i\infty}dt_1~\Gamma(t_1-t+\tfrac{1}{2})
\zeta(2t_1-2t+1)\Gamma(t_1)\left(\frac{ML_0}{2}\right)^{-2t_1}
\Biggr\}\notag\\
&\hspace{5mm}\times 
\left(\frac{\pi}{L_0}\right)^{-2t}{\tilde S}^{(n)}(t+\tfrac{D-1}{2}; L_{i_1},\cdots, L_{i_n}).
\label{4.5}
\end{align}
We deform the integration path with respect to $t_1$ in Eq.(\ref{4.5}) in such a way that it encloses all the poles in the integrand and 
perform the $t_1$ integration by the residue theorem. Among the 
poles $t_1=t-\half -\ell~(\ell=1,2,\cdots)$ of $\Gamma(t_1-t+\tfrac{1}{2})$, only the pole
$t_1=t-\tfrac{1}{2}$ is relevant because of the property $\zeta(-2\ell)=0~(\ell=1,2,\cdots)$, which always follows from the 
combination $\Gamma(t_1-t+\tfrac{1}{2})\zeta(2t_1-2t+1)$ and is frequently used throughout 
our discussions. The residue integral from the pole $t_1=t-\tfrac{1}{2}$ with $\zeta(0)=-1/2$ cancels the second term in Eq.(4.5) and 
the integral from the pole $t_1=t$ of $\zeta(2t_1-2t+1)$ with $\Gamma(1/2)=\sqrt{\pi}$ does the first terms in Eq.(4.5). We are left with the
contribution from the pole $t_1=-\bar n~(\bar n=0,1,\cdots)$ of $\Gamma(t_1)$, so that we find that 
the second term in Eq.(\ref{4.1}) becomes 
\begin{align}
&-\frac{2^{n+1}}{\pi^{\frac{D}{2}}L_0}\frac{1}{4\pi i}\int_{\bar c -i\infty}^{\bar c+i\infty}dt~
\sum_{\bar n=0}^{\infty}\frac{(-1)^{\bar n}}{\bar n!}\Gamma(-\bar n -t+\tfrac{1}{2})
\zeta(-2\bar n -2t +1)\left(\frac{ML_0}{2}\right)^{2\bar n}
\notag\\
&\hspace{5mm}\times \left(\frac{1}{L_0}\right)^{-2t}{\tilde S}^{(n)}(t+\tfrac{D-1}{2}; L_{i_1},\cdots, L_{i_n}).
\label{4.6}
\end{align}

Even though there would appear the poles in the multiple mode summations ${\tilde S}^{(n)}(t+\tfrac{D-1}{2};L_{i_1},\cdots, L_{i_n})$ 
that contribute to the $t$ integration in Eq.(\ref{4.6}) by the residue theorem, it does not affect the power of the mass 
squared $(M^2)^{\bar n}$ in Eq.(\ref{4.6}). Thus, Eq.(\ref{4.6}) has only positive integer powers of the mass 
squared $M^2$, {\it i.e.} $(M^2)^{\bar n}$. This observation implies that 
only the first term in Eq.(\ref{4.1}) potentially can have the nonanalytic terms, so that from Eq.(\ref{3.3}), we obtain 
the important relation
\begin{align}
(F_{L_{i_1},\cdots, L_{i_n}}^{(n)D}(M)+F_{L_0, L_{i_1},\cdots, L_{i_n}}^{(n+1)D}(M))\Big|_{\rm n.a.}
&=
\frac{1}{L_0}\sum_{n_0=-\infty}^{\infty}F_{L_{i_1},\cdots, L_{i_n}}^{(n)D-1}(M_{(0)})\Big|_{\rm n.a.}
\notag\\
&
=
\frac{1}{L_0}F_{L_{i_1},\cdots, L_{i_n}}^{(n)D-1}(M)\Big|_{\rm n.a.}.
\label{4.7}
\end{align}
The abbreviation denoted by ``{\rm n.a.}'' in Eq.(\ref{4.7}) means {\it nonanalytic terms}, which is used 
throughout the paper. It should be stressed that only the zero mode $(n_0=0)$ in $M_{(0)}$ is relevant for the existence of the 
nonanalytic terms in the mode recombination formula (\ref{3.3}). The relation (\ref{4.7}) is crucial for the continuing discussion given below.

The relevant part for the nonanalytic terms in the effective potential (\ref{3.8}) is given by 
replacing $M_{(0)}$ by $M$ without the summation because of Eq.(\ref{4.7}).
Considering the discussion led to Eq.(\ref{4.7}), we understand that the relevant part for 
the nonanalytic terms in Eq.(\ref{3.9}) is given by replacing the $M_{(0)}, M_{(0,1)}$ by $M, M_{(1)}$, respectively and dropping the 
$n_0$ summation. By repeating the same discussion as above and applying it to the effective 
potential (\ref{3.11}), the relevant part of the nonanalytic terms 
in the effective potential is given by picking up only the zero modes in $M_{(0,1,\cdots,k-1)}~(k=1,2,\cdots,p)$, that is, 
\begin{align}
V_{\rm eff}\big|_{\rm n.a.}
&=F^{(0)D}(M)\Big|_{\rm n.a.}+\sum_{k=0}^p\left(\frac{1}{\prod_{i=0}^{k-1}L_i}\right)F_{L_k}^{(1)D-k}(M)\Big|_{\rm n.a.}.
\label{4.8}
\end{align}
%
%
%
%This is the simplified expression for the effective potential, by which we calculate the nonanalytic term.
%
%We stress that 
%
It  turns out that the relevant part of the nonanalytic terms in the effective potential 
drastically reduces to the simple expression (\ref{4.8}) and is given by the contribution of the {\it single} mode
summation of the winding mode $m_k~(k=0,\cdots, p)$ associated with the $S_k^1$. 
This is quite different from that given in the previous paper where we have analyzed the multiple mode 
summations in order to obtain the nonanalytic terms. 
It should be emphasized that Eq.(\ref{4.8}) is obtained as the consequence of taking account of the 
zero mode $(n_0=\cdots =n_{p}=0)$ alone in the Matsubara and Kaluza-Klein modes.

Let us confirm the results obtained in the previous paper by using Eq.(\ref{4.8}). We will soon recognize that 
the calculations based on Eq.(\ref{4.8}) are easier and more transparent compared with that in the previous paper.
We first study the nonanalytic terms of $F_{L_k}^{(1)D-k}(M)$ in Eq.(\ref{4.8}). From Eqs.(\ref{2.16}) and (\ref{3.18}), one has (remember 
$\eta_0=0, {\cal N}=1, \eta_k=0$ in this section)
\begin{align}
F_{L_k}^{(1)D-k}(M)&=\frac{-2}{(2\pi)^{\frac{D-k}{2}}}\sum_{m_k=1}^{\infty}\left(\frac{M^2}{(m_kL_k)^2}\right)^{\frac{D-k}{4}}
K_{\frac{D-k}{2}}(m_kML_k)\notag\\
&=\frac{-2}{(2\pi)^{\frac{D-k}{2}}}\left(\frac{M^2}{2}\right)^{\frac{D-k}{2}}
\frac{1}{4\pi i} \int_{c-i\infty}^{c+i\infty}dt~\Gamma(t-\tfrac{D-k}{2})\Gamma(t)\zeta(2t)\left(\frac{ML_k}{2}\right)^{-2t},
\label{4.9}
\end{align}
where we have used the definition of the zeta function $\zeta(z)=\sum_{m=1}^{\infty}m^{-z}$. 
%
%
%We deform the integration path in such a way that it encloses all the poles 
%of the integrand and evaluate the $t$ integration by the residue theorem. 
%
%
Since the scale dependence on 
$M$ in Eq.(\ref{4.9}) is $(M^2)^{\frac{D-k}{2}-t}$, the poles $t=\tfrac{D-k}{2}-\ell~(\ell=0,1,\cdots)$ of $\Gamma(t-\frac{D-k}{2})$
yields analytic terms $(M^2)^{\ell}$, which are outside of our interest.
Only the pole $t=0$ among the poles $t=-m~(m=0,1,\cdots,)$ of $\Gamma(t)$ contributes to the residue integration
because of the property $\zeta(-2m)=0~(m=1,2,\cdots)$. 
Hence, there are two poles, $t=0$ of $\Gamma(t)$ and $t=\half$ of $\zeta(2t)$  
that contribute to the residue integration in Eq.(\ref{4.9}) for obtaining the nonanalytic terms\footnote{The $\log M $ is another 
type of the nonanalytic term in the effective potential, which is obtained by the residue integration for the 
double pole of the integrand in Eq.(\ref{4.9}). We do not study such the term in this paper.}. 
%
%And the (2.3) also possesses the $\log M$ term for $D=$ even. 
%

%
%Since the nonanalytic term has the form of any positive odd power of $M$, which pole, $t=0$ or $t=\half$ produces the nonanalytic term 
%depends on even/odd $D$ and $k$. 
%
Since the nonanalytic terms are given by positive odd integer powers of $M$, whether the
pole of $t=0$ or $t=1/2$ can produce a nonanalytic term depends on whether $D$ and $k$
are even  or odd. For $(D, k)=({\rm even}, {\rm even})$ or $({\rm odd}, {\rm odd})$, the nonanalytic 
terms in Eq.(\ref{4.9}) is given by the pole $t=\half$ as
\begin{align}
F_{L_k}^{(1)D-k}(M)\Big|_{\rm n.a.}&=\frac{(-1)(-1)^{\frac{D-k}{2}}}{2^{\frac{D-k}{2}}\pi^{\frac{D-k-2}{2}}(D-k-1)!!}\frac{M^{D-k-1}}{L_k},
\label{4.10}
\end{align}
while for $(D, k)=({\rm even}, {\rm odd})$ or $({\rm odd}, {\rm even})$, the pole $t=0$ yields the nonanalytic term
\begin{align}
F_{L_k}^{(1)D-k}(M)\Big|_{\rm n.a.}&=\frac{(-1)^{\frac{D-k+1}{2}}}{2^{\frac{D-k+1}{2}}\pi^{\frac{D-k-1}{2}}(D-k)!!}M^{D-k}.
\label{4.11}
\end{align}
Here, we have used 
\begin{align}
\Gamma(\tfrac{1-D}{2})=\frac{(-1)^{\frac{D}{2}}2^{\frac{D}{2}}}{(D-1)!!}\sqrt{\pi}\quad {\rm for}\quad D={\rm even}.
\label{4.12}
\end{align}

We are ready to calculate the nonanalytic terms in the effective 
potential based on Eq.(\ref{4.8}) by using Eqs.(\ref{4.10}) and (\ref{4.11}). It is convenient to study the terms for each
case of even/odd $D$ and $p+1$. Let us remind that $p+1$ is the total number of $S^1$.
%
%
%We sometimes omit the abbreviation ``n.a.'' in equations hereafter, but we should keep in mind that we are treating the 
%nonanalytic term. 
%
%
%
%
\subsection{$(D, p+1)=$ (even, odd)}
Let us first introduce
\begin{align}
A_{2n+1}&\equiv \frac{1}{\prod_{i=0}^{2n}L_i}F_{L_{2n+1}}^{(1)D-(2n+1)}(M)\Big|_{\rm n.a.}=
\frac{(-1)^{\frac{D-2n}{2}}}{2^{\frac{D-2n}{2}}\pi^{\frac{D-2n-2}{2}}(D-(2n+1))!!}\frac{M^{D-(2n+1)}}{L_0\cdots L_{2n}},
\label{4.13}
\\
B_{2n}&\equiv \frac{1}{\prod_{i=0}^{2n-1}L_i}F_{L_{2n}}^{(1)D-2n}(M)\Big|_{\rm n.a.}=
\frac{(-1)(-1)^{\frac{D-2n}{2}}}{2^{\frac{D-2n}{2}}\pi^{\frac{D-2n-2}{2}}(D-2n-1)!!}\frac{M^{D-2n-1}}{L_0\cdots L_{2n}},
\label{4.14}
\end{align}
where we have used Eq.(\ref{4.11})  (Eq.(\ref{4.10})) in the equality of Eq.(\ref{4.13}) (Eq.(\ref{4.14})). One immediately observes that
\begin{align}
A_{2n+1}=-B_{2n}\quad {\rm for}\quad n=0,1,\cdots.
\label{4.15}
\end{align} 
Let us note that the definition for $B_0$ in Eq.(\ref{4.14}) is consistently equal to $F_{L_0}^{(1)D}(M)$ by 
setting $\prod_{i=0}^{2n-1}L_i\big|_{n=0}=1$, which originally corresponds to the $k=0$ term in Eq.(\ref{3.11}).

In terms of Eqs.(\ref{4.13}) and (\ref{4.14}), the nonanalytic term in Eq.(\ref{4.8}) is given by
\begin{align}
V_{\rm eff}\big|_{\rm n.a.}=F^{(0)D}(M)\big|_{\rm n.a.}+\sum_{n=0}^{\tfrac{p}{2}-1}(A_{2n+1}+B_{2n})+B_p=B_p,
\label{4.16}
\end{align}
where we have used Eq.(\ref{4.15}) and the fact that from Eq.(\ref{2.13}), the $F^{(0)D}(M)$ does not possess any nonanalytic 
term for $D=$ even. The $B_p$ is given by Eq.(\ref{4.14}) with $n=\frac{p}{2}$ to yield, by noting 
that $(-1)(-1)^{\frac{D-p}{2}}=(-1)^{\frac{D+p+2}{2}}$ for $p=$ even, 
\begin{align}
V_{\rm eff}\big|_{\rm n.a.}=B_p=\frac{(-1)^{\frac{D+p+2}{2}}}{2^{\frac{D-p}{2}}\pi^{\frac{D-p-2}{2}}(D-(p+1))!!}
\frac{M^{D-(p+1)}}{L_0L_1\cdots L_p}.
\label{4.17}
\end{align}
%
%We have correctly reproduced the result in the previous paper for the case.
%
%
%
%
\subsection{$(D, p+1)=$ (even, even)}
In this case, the nonanalytic term in Eq.(\ref{4.8}) is written by using Eqs.(\ref{4.13}) and (\ref{4.14}) as 
\begin{align}
V_{\rm eff}\big|_{\rm n.a.}=F^{(0)D}(M)\big|_{\rm n.a.}+\sum_{n=0}^{\tfrac{p-1}{2}}(A_{2n+1}+B_{2n})=0,
\label{4.18}
\end{align}
where we have used Eq.(\ref{4.15}) in the last equality. The effective potential does not have any nonanalytic term in this case.
\subsection{$(D, p+1)=$ (odd, odd)}
For the case of $D=$ odd, from Eqs.(\ref{4.10}) and (\ref{4.11}), it is convenient to introduce
\begin{align}
C_{2n-1}&\equiv \frac{1}{\prod_{i=0}^{2n-2}L_i}F_{L_{2n-1}}^{(1)D-(2n-1)}(M)\Big|_{\rm n.a.}=
\frac{(-1)(-1)^{\frac{D-(2n-1)}{2}}}{2^{\frac{D-(2n-1)}{2}}\pi^{\frac{D-2n-1}{2}}(D-2n)!!}\frac{M^{D-2n}}{L_0\cdots L_{2n-1}},
\label{4.19}\\
D_{2n}&\equiv \frac{1}{\prod_{i=0}^{2n-1}L_i}F_{L_{2n}}^{(1)D-2n}(M)\Big|_{\rm n.a.}=
\frac{(-1)^{\frac{D-2n+1}{2}}}{2^{\frac{D-2n+1}{2}}\pi^{\frac{D-2n-1}{2}}(D-2n)!!}\frac{M^{D-2n}}{L_0\cdots L_{2n-1}},
\label{4.20}
\end{align}
respectively. We see that the relation 
\begin{align}
C_{2n-1}=-D_{2n}\quad {\rm for}\quad n=0,1,\cdots
\label{4.21}
\end{align}
holds. It should be understood that in Eq.(\ref{4.21}) we have defined $C_{2n-1}|_{n=0}\equiv F^{(0)D}(M)$ 
and $D_{2n}|_{n=0}\equiv F_{L_0}^{(1)D}(M)$.
One immediately confirms that the $C_{-1}=-D_0$ is satisfied  by using the explicit expressions for
$F^{(0)D}(M)$ in Eq.(\ref{2.14}) and $F_{L_0}^{(1)D}(M)$\cite{sakatake2022} in Eq.(4.11) for $D=$ odd.
%
%
% that corresponds to $k=0$ term in (3.11). 
%These are consistent with the explicit expressions for $F^{(0)D}(M)$ (2.14) and $ F_{L_0}^{(1)D}(M)$\cite{sakatake2022} 
%if we set $\prod_{i=0}^{2n-1}L_i|_{n=0}=1$ in (4.18) and (4.19). 
%

In terms of Eqs.(\ref{4.19}) and (\ref{4.20}), the nonanalytic part of the effective potential in Eq.(\ref{4.8}) is calculated as 
\begin{align}
V_{\rm eff}\big|_{\rm n.a.}=\sum_{n=0}^{\frac{p}{2}}(C_{2n-1}+D_{2n})=0,
\label{4.22}
\end{align}
thanks to Eq.(\ref{4.21}).
\subsection{$(D, p+1)=$ (odd, even)}
In this case, the nonanalytic term in Eq.(\ref{4.8}) is obtained as
\begin{align}
V_{\rm eff}\big|_{\rm n.a.}=\sum_{n=0}^{\frac{p-1}{2}}(C_{2n-1}+D_{2n})+C_p=C_p,
\label{4.23}
\end{align}
where we have used Eq.(\ref{4.21}). The $C_p$ is given by Eq.(\ref{4.19}) with $n=\frac{p+1}{2}$ to yield
\begin{align}
V_{\rm eff}\big|_{\rm n.a.}=C_p=
\frac{(-1)^{\frac{D+p}{2}}}{2^{\frac{D-p}{2}}\pi^{\frac{D-p-2}{2}}(D-(p+1))!!}
\frac{M^{D-(p+1)}}{L_0L_1\cdots L_p},
\label{4.24}
\end{align}
where we have used the fact $(-1)(-1)^{\frac{D-p}{2}}=(-1)^{\frac{D+p}{2}}$ for $p=$ odd.

The results (\ref{4.17}), (\ref{4.18}), (\ref{4.22}) and (\ref{4.24}) are found to exactly agree with those given in the previous paper, as they should be.
Even though they have already been obtained in the previous paper, our formalism developed in this paper is easier and 
more transparent to derive the nonanalytic terms of the effective potential, and furthermore makes it possible to
analyze the case of the fermion field, as we will see in the next section.
In addition to it, a new insight on the nonanalytic term is obtained. 
The nonanalytic term Eq.(\ref{4.17}) (or Eq.(\ref{4.24})) depends 
on each scale, $L_0,\cdots, L_p$ of all the $S^1$'s. This may reflect the fact that only the zero modes in 
the Matsubara and Kaluza-Klein modes are relevant for the existence of the nonanalytic terms, as shown in Eq.(\ref{4.8}).
%
%that the nonanalytic term exists only when 
%there is the zero mode associated with each of all the $S^1$'s. 
%
%This suggestion may 
%
%
%
%
%
%
%
%%%%%%%%%%%%%%%%%%%%%%%%%%%%%%%%%%%%%%
\section{Nonanalytic terms for fermion field}
%%%%%%%%%%%%%%%%%%%%%%%%%%%%%%%%%%%%%%
%
%
%
Let us study the nonanalytic terms in the effective potential for the case of the fermion 
field $(f=1, \eta_0=\half)$ satisfying arbitrary boundary condition for the spatial $S_i^1~(i=1,2,\cdots, p)$ direction. The 
effective potential (\ref{3.15}) rather than Eq.(\ref{3.11}) is appropriate to discuss the terms in this case.

We first consider the $n\geq 1$ term, the second one in Eq.(\ref{3.15}), whose integral form on the 
complex plane is given from Eq.(\ref{3.19}) by 
\begin{align}
&\frac{1}{L_0}\sum_{n_0=-\infty}^{\infty}F_{L_{i_1},\cdots, L_{i_n}}^{(n)D-1}(M_{(0)})\notag\\
&=\frac{1}{L_0}{\cal N}\frac{2^n}{\pi^{\frac{D-1}{2}}}\frac{1}{4\pi i}
\int_{\bar c-i\infty}^{\bar c+i\infty}dt~\Gamma(t)\sum_{n_0=-\infty}^{\infty}
\Biggl\{
\left(n_0+\half\right)^2+\left(\frac{ML_0}{2\pi}\right)^2
\Biggr\}^{-t}\notag\\
&\hspace{5mm}\times 
\left(\frac{\pi}{L_0}\right)^{-2t}S^{(n)}(t+\tfrac{D-1}{2};L_{i_1},\cdots, L_{i_n}),
\label{5.1}
\end{align}
where we have changed the variable $\bar t=t-\tfrac{D-1}{2}$ and have denoted $\bar t$ by $t$ again.
%
%
%where we have not written the irrelevant summations in (3.15) for the discussion.
%we have defined
%\begin{align}
%{\tilde S}^{(n)}(t+\tfrac{D-1}{2}; L_{i_1},\cdots, L_{i_n})&\equiv \Gamma(t)\sum_{i_1=1}^{\infty}\cdots\sum_{i_n=1}^{\infty}
%\left\{(m_{i_1}L_{i_1})^2+\cdots +(m_{i_n}L_{i_n})^2
%\right\}^{-t-\frac{D-1}{2}}\notag\\
%&\hspace{5mm}\times \cos(2\pi m_{i_1}\eta_{i_1})\cdots \cos(2\pi m_{i_n}\eta_{i_n})
%\end{align} 
%we have used the (3.4) with $\eta_0=\frac{1}{2}$.
%
%
%

One needs to develop the analytical extension for the mode summation with respect to $n_0$ in Eq.(\ref{5.1}). To this end, we 
employ the formula (2.8) and the Poisson summation (\ref{2.9}). Then, we find 
\begin{align}
\Gamma(t)\sum_{n_0=-\infty}^{\infty}
\Biggl\{
\left(n_0+\half\right)^2+z^2\Biggr\}^{-t}=
\sqrt{\pi}\sum_{m_0=-\infty}^{\infty}\int_0^{\infty}d{\bar t}~{\bar t}\,^{(t-\half)-1}~
\e^{-\frac{(2\pi m_0)^2}{4\bar t}+\pi im_0-z^2\bar t}.
\label{5.2}
\end{align}
By separating $m_0=0$ and $m_0\neq 0$ modes in Eq.(\ref{5.2}) and by using Eqs.(\ref{2.8}) and (\ref{2.15}), the 
right-hand side of Eq.(\ref{5.2}) becomes
\begin{align}
%
%&\sum_{n_0=-\infty}^{\infty}
%\Biggl\{
%\left(n_0+\half\right)^2+c^2\Biggr\}^{-t}\notag\\
%&\hspace{5mm}=
%
&\sqrt{\pi}\frac{\Gamma(t-\half)}{z^{2(t-\half)}}
+4\sqrt{\pi}\left(\frac{\pi}{z}\right)^{t-\half}\sum_{m_0=1}^{\infty}(-1)^{m_0}
m_0^{t-\half}K_{t-\half}(2\pi m_0 z)\notag\\
&\hspace{5mm}
=\sqrt{\pi}\frac{\Gamma(t-\half)}{z^{2(t-\half)}}
+4\sqrt{\pi}\left(\frac{\pi}{z}\right)^{t-\half}
\Biggl\{-\sum_{m_0=1}^{\infty}m_0^{t-\half}K_{t-\half}(2\pi m_0 z)+2\sum_{m_0=1}^{\infty}(2m_0)^{t-\half}K_{t-\half}(4\pi m_0 z)
\Biggr\},
\label{5.3}
\end{align}
%
%Here, we have arranged the mode summation with respect to $m_0$ so as not to have the factor $(-1)^{m_0}$ for 
%latter convenience. 
%
where we have used the property $K_{-\nu}(z)=K_{\nu}(z)$ and the formula 
\begin{align}
\sum_{m_0=1}^{\infty}(-1)^{m_0}f(m_0)=-\sum_{m_0=1}^{\infty}f(m_0)+2\sum_{m_0=1}^{\infty}f(2m_0)
\label{5.4}
\end{align}
for later convenience. We further recast Eq.(\ref{5.3}) into the integral form on the complex plane by Eq.(\ref{3.18}). Then, we obtain
\begin{align}
&\Gamma(t)\sum_{n_0=-\infty}^{\infty}
\Biggl\{
\left(n_0+\half\right)^2+z^2\Biggr\}^{-t}\notag\\
&\hspace{5mm}
=\sqrt{\pi}\frac{\Gamma(t-\half)}{z^{2(t-\half)}}
-\frac{4\pi^{2t}}{\sqrt{\pi}}\frac{1}{4\pi i}\int_{c_1-i\infty}^{c_1+i\infty}dt_1~\Gamma(t_1-t+\tfrac{1}{2})\zeta(2t_1-2t+1)\Gamma(t_1)(\pi z)^{-2t_1}\notag\\
&\hspace{3.2cm}
+\frac{4(2\pi)^{2t}}{\sqrt{\pi}}\frac{1}{4\pi i}\int_{c_1-i\infty}^{c_1+i\infty}dt_1~\Gamma(t_1-t+\tfrac{1}{2})\zeta(2t_1-2t+1)\Gamma(t_1)(2\pi z)^{-2t_1}.
\label{5.5}
\end{align}
This is an analytical extension for the mode summation with respect to $n_0$ in Eq.(\ref{5.1}).
Eq.(\ref{5.5}) does not have the term corresponding to the first term in Eq.(\ref{4.4}) because of the lack of the 
zero mode due to the antiperiodic boundary condition for the fermion field in the Euclidean time direction.

By setting $z=\frac{ML_0}{2\pi}$ and inserting Eq.(\ref{5.5}) into Eq.(\ref{5.1}), we have
\begin{align}
&\frac{1}{L_0}\sum_{n_0=-\infty}^{\infty}F_{L_{i_1},\cdots, L_{i_n}}^{(n)D-1}(M_{(0)})\notag\\
&=\frac{1}{L_0}{\cal N}\frac{2^n}{\pi^{\frac{D-1}{2}}}\frac{1}{4\pi i}\int_{c-i\infty}^{c+i\infty}dt~\Biggl\{
\sqrt{\pi}\Gamma(t-\tfrac{1}{2})\left(\frac{ML_0}{2\pi}\right)^{-2(t-\half)}\notag\\
&\hspace{5mm}
-\frac{4\pi^{2t}}{\sqrt{\pi}}\frac{1}{4\pi i}\int_{c_1-i\infty}^{c_1+i\infty}dt_1~\Gamma(t_1-t+\tfrac{1}{2})\zeta(2t_1-2t+1)\Gamma(t_1)
\left(\frac{ML_0}{2}\right)^{-2t_1}
\notag\\
&\hspace{5mm}
+\frac{4(2\pi)^{2t}}{\sqrt{\pi}}\frac{1}{4\pi i}\int_{c_1-i\infty}^{c_1+i\infty}dt_1~\Gamma(t_1-t+\tfrac{1}{2})\zeta(2t_1-2t+1)\Gamma(t_1)
\left(ML_0\right)^{-2t_1}
\Biggr\}\notag\\
&\hspace{5mm}\times
\left(\frac{\pi}{L_0}\right)^{-2t}S^{(n)}(t+\tfrac{D-1}{2}; L_{i_1},\cdots, L_{i_n}).
\label{5.6}
\end{align}
We perform the residue integration with respect to $t_1$ by deforming the integration path in such a way that it 
encloses all the poles in the integrand. Among the poles $t_1=t-\half -\ell~(\ell=0,1,\cdots)$ of $\Gamma(t_1-t+\half)$, only 
the pole $t_1=t-\half$ is relevant to the residue integration thanks to the property $\zeta(-2\ell)=0~(\ell=1,2,\cdots)$. 
In addition to it, the poles $t_1=t$ of $\zeta(2t_1-2t +1)$ and $t_1=-\bar n~(\bar n=0,1,\cdots)$
of $\Gamma(t_1)$ contribute to the $t_1$ integration by the residue theorem.
It is easy to see that the contributions from the residue integration of the 
pole $t_1=t-\half$ in the second and third terms of Eq.(\ref{5.6}) cancel the first term in Eq.(\ref{5.6}) and that the contributions 
from the residue integration of the pole $t_1=t$ of $\zeta(2t_1-2t+1)$ 
in the second and the third terms of Eq.(\ref{5.6}) are canceled each other. Thus, what is 
left is the contribution of the residue integration of the pole $t_1=-\bar n~(\bar n=0,1,\cdots)$ of $\Gamma(t_1)$ alone, which is given by
\begin{align}
&\frac{1}{L_0}\sum_{n_0=-\infty}^{\infty}F_{L_{i_1},\cdots, L_{i_n}}^{(n)D-1}(M_{(0)})\notag\\
\hspace{5mm}
&=\frac{1}{L_0}{\cal N}\frac{2^n}{\pi^{\frac{D-1}{2}}}\frac{1}{4\pi i}
\int_{c-i\infty}^{c+i\infty}dt~\sum_{\bar n=0}^{\infty}
\Biggl\{
-\frac{2}{\sqrt{\pi}}\frac{(-1)^{\bar n}}{\bar n!}\Gamma(-\bar n-t+\tfrac{1}{2})\zeta(-2\bar n -2t+1)
\left(\frac{ML_0}{2}\right)^{2\bar n}L_0^{2t}\notag\\
&\hspace{1cm}+
\frac{2}{\sqrt{\pi}}\frac{(-1)^{\bar n}}{\bar n!}\Gamma(-\bar n-t+\tfrac{1}{2})\zeta(-2\bar n-2t+1)
\left(ML_0\right)^{2\bar n}(2L_0)^{2t}
\Biggr\}S^{(n)}(t+\tfrac{D-1}{2};L_{i_1},\cdots, L_{i_n}).
\label{5.7}
\end{align}

One observes that the dependence of the power on $M$ in Eq.(\ref{5.7}) is given by the positive integer power of the mass 
squared $(M^2)^{\bar n}$, so that there exists no nonanalytic term in the second term in Eq.(\ref{3.15})
\begin{align}
\sum_{n=1}^{p}~\sum_{1\leq i_1<i_2<\cdots < i_n\leq p}
\frac{1}{L_0}\sum_{n_0=-\infty}^{\infty}F_{L_{i_1},\cdots, L_{i_n}}^{(n)D-1}(M_{(0)})\Big|_{\rm n.a.}=0.
\label{5.8}
\end{align}
We note that the multiple mode summations in $S^{(n)}(t+\frac{D-1}{2};L_{i_1},\cdots,L_{i_n})$ could produce poles
that contribute to the $t$ integration in Eq.(\ref{5.7}), but they do not change the power of $(M^2)^{\bar n}$ because of the nonexistence 
of $M$ in the summations. It is important to note that Eq.(\ref{5.7}) holds irrespective of the boundary
condition for the spatial $S_i^1~(i=1,2,\cdots, p)$ direction and thus, so does Eq.(\ref{5.8}).

Let us next study the $n=0$ term, the first one in Eq.(\ref{3.15}), which is given by Eq.(\ref{3.17}) with $f=0$ and $\eta_0=\half$. 
We make use of the analytical extension (\ref{5.4}) in order to calculate the mode summation 
with respect $n_0$ in Eq.(\ref{3.17}). By putting $t=-\frac{D-1}{2}, z=\frac{ML_0}{2\pi}$ in Eq.(\ref{5.5}), the relevant 
part of Eq.(\ref{3.17}) becomes 
\begin{align}
&\Gamma(-\tfrac{D-1}{2})\sum_{n_0=-\infty}^{\infty}
\Biggl\{
\left(n_0+\half\right)^2+\left(\frac{ML_0}{2\pi}\right)^2\Biggr\}^{\frac{D-1}{2}}\notag\\
&
=\sqrt{\pi}\Gamma(-\tfrac{D}{2})\left(\frac{ML_0}{2\pi}\right)^D\notag\\
&\hspace{5mm}
-\frac{4\pi^{-(D-1)}}{\sqrt{\pi}}\frac{1}{4\pi i}\int_{c_1-i\infty}^{c_1+i\infty}dt_1~\Gamma(t_1+\tfrac{D}{2})\zeta(2t_1+D)\Gamma(t_1)
\left(\frac{ML_0}{2}\right)^{-2t_1}\notag\\
&\hspace{5mm}
+\frac{4(2\pi)^{-(D-1)}}{\sqrt{\pi}}\frac{1}{4\pi i}\int_{c_1-i\infty}^{c_1+i\infty}dt_1~\Gamma(t_1+\tfrac{D}{2})\zeta(2t_1+D)\Gamma(t_1)
\left(ML_0\right)^{-2t_1}.
\label{5.9}
\end{align}
We again evaluate the $t_1$ integration by the residue theorem by deforming the integration path 
in such a way that it encloses all the poles in the integrand. The nonanalytic 
terms\footnote{We do not consider the nonanalytic terms of the type, $\log M$ in this paper, which comes from the double pole
in the integrand, as mentioned in the footnote of the section $4$.} in the second and the third 
terms in Eq.(\ref{5.9}) are given by the residue integration of the poles that 
have the minus half odd integer values of $t_1$, as seen from the scale dependence 
on $M$ in Eq.(\ref{5.9}).

For $D=$ even, the relevant pole is $t_1=-\tfrac{D-1}{2}$ of $\zeta(2t_1+D)$, from which the second 
and the third terms in Eq.(5.9) yield
\begin{align}
-\frac{4\sqrt{\pi}}{\sqrt{\pi}2^2}\Gamma(-\tfrac{D-1}{2})\left(\frac{ML_0}{2\pi}\right)^{D-1}+
\frac{4\sqrt{\pi}}{\sqrt{\pi}2^2}\Gamma(-\tfrac{D-1}{2})\left(\frac{ML_0}{2\pi}\right)^{D-1}=0.
\label{5.10}
\end{align}
The first term in Eq.(\ref{5.9}) is analytic for $D=$ even, so that there is no nonanalytic 
term in Eq.(\ref{5.9}) for $D=$ even.

On the other hand, for $D=$ odd, the relevant pole for the nonanalytic terms is given only by the 
pole $t_1=-\tfrac{D}{2}$ among the poles $t_1=-\tfrac{D}{2}-\ell~(\ell=0,1,\cdots)$ of $\Gamma(t_1+\tfrac{D}{2})$ because of 
the property $\zeta(-2\ell)=0~(\ell=1,2,\cdots)$. Then, the second and the third terms in Eq.(\ref{5.9}) lead to
\begin{align}
\sqrt{\pi}\Gamma(-\tfrac{D}{2})\left(\frac{ML_0}{2\pi}\right)^D - 2\sqrt{\pi}\Gamma(-\tfrac{D}{2})\left(\frac{ML_0}{2\pi}\right)^D
=-\sqrt{\pi}\Gamma(-\tfrac{D}{2})\left(\frac{ML_0}{2\pi}\right)^D,
\label{5.11}
\end{align}
which cancels the first term in Eq.(\ref{5.9}). Thus, Eq.(\ref{5.9}) has no nonanalytic term for $D=$ odd.
%
%
%
%\begin{align}
%\Gamma(-\tfrac{D-1}{2})\sum_{n_0=-\infty}^{\infty}
%\Biggl\{
%\left(n_0+\half\right)^2+\left(\frac{ML_0}{2\pi}\right)^2\Biggr\}^{\frac{D-1}{2}}\Bigg|_{\rm n.a.}=0,
%\end{align}
%
%

We have shown that the first term in Eq.(\ref{3.15}) does not possess the nonanalytic term
\begin{align}
\frac{1}{L_0}\sum_{n_0=-\infty}^{\infty}F^{(0)D-1}(M_{(0)})\Big|_{\rm n.a.}=0.
\label{5.12}
\end{align}
We arrive at the important conclusion from Eqs.(\ref{5.8}) and (\ref{5.12}) that the effective 
potential (\ref{3.16}) has no nonanalytic term for the case of the fermion
\begin{align}
V_{\rm eff}\big|_{\rm n.a.}=
\sum_{n=0}^{p}~\sum_{1\leq i_1<i_2<\cdots < i_n\leq p}\frac{1}{L_0}\sum_{n_0=-\infty}^{\infty}
F_{L_{i_1},\cdots, L_{i_n}}^{(n)D-1}(M_{(0)})\Big|_{\rm n.a.}=0.
\label{5.13}
\end{align}
It should be emphasized that Eq.(\ref{5.13}) holds irrespective of the boundary condition 
for the spatial $S_i^1~(i=1,2,\cdots, p)$ direction. 
%
%
%
%
%\begin{align}
%\frac{1}{L_0}\sum_{n_0=-\infty}^{\infty}F_{L_{i_1},\cdots, L_{i_n}}^{(n)D-1}(M_{(0)})\Big|_{\rm n.a.}=0.
%\end{align} 
%Since the effective potential is given by (3.16), we immediately obtain that
%
%
%The nonzero Matsubara mass removal of the zero mode for the Euclidean time direction 
%
%
%the (5.7) holds irrespective of the boundary condition for the spatial $S^1$ direction, 
%the result (5.12) holds for the case of the fermion field satisfying arbitrary boundary condition 
%for the spatial $S^1$ direction.
%
%
%%%%%%%%%%%%
\section{Nonanalytic terms for scalar field with antiperiodic boundary condition}
%%%%%%%%%%
Taking account of the result obtained in the previous section, we can also study the case of the scalar field with some of the 
boundary condition for the spatial $S_i^1~(i=1,\cdots, p)$ direction being antiperiodic. In this case, by regarding the direction that has the 
antiperiodic boundary condition as the Euclidean time direction, the effective potential essentially has the same with 
that of the case for the fermion field. Hence, we conclude that there is no nonanalytic term in the effective potential
if the scalar field satisfies at least one antiperiodic boundary condition for the spatial $S_i^1$ direction.
%
%
%
%%%%%%%%%%%%%%%%%%%%%%%%%%%%%%%%%%%%%%
\section{Conclusions and discussions}
%%%%%%%%%%%%%%%%%%%%%%%%%%%%%%%%%%%%%%
%
%
%
We have studied the nonanalytic terms, which cannot be written in the form of any positive integer power of 
field-dependent mass squared, in the effective potential at finite temperature in one-loop approximation 
for the fermion and scalar fields
on the $D$-dimensional spacetime, $S_{\tau}^1\times R^{D-(p+1)}\times \prod_{i=1}^pS_i^1$.
In doing it, we have developed the new formula called the mode recombination formula (\ref{3.3}), which holds 
irrespective of whether the field is a fermion or a scalar and of the boundary condition
for the spatial $S_i^1$ direction. The effective potential has been recast into 
the new forms (\ref{3.11}) and (\ref{3.16}) by using 
the formula, which is convenient to study the nonanalytic terms for both cases of the fermion and the scalar.

We have clarified the importance of the zero mode in 
the Kaluza-Klein mode for the existence of the nonanalytic terms through the
mode recombination formula. This has drastically simplified
the relevant part of the effective potential for calculating the nonanalytic terms for the case of the scalar field 
satisfying the periodic boundary condition for the spatial $S_i^1$ direction. 
The effective potential (\ref{4.8}) is given in terms of the single mode summation of the winding 
mode for each $S^1$ and the integral form for the potential on the complex 
plane is easy to perform the residue integration. We have correctly reproduced the 
nonanalytic term (\ref{4.17}) (and (\ref{4.24})) in easier and more transparent way.

The mode recombination formula has also provided the convenient form for studying
the nonanalytic terms in the effective potential for the case of the fermion field satisfying arbitrary 
boundary conditions for the spatial $S_i^1~(i=1,\cdots, p)$ directions. The antiperiodicity for the 
Euclidean time direction for the fermion has resulted the quite different pole structure of the analytical 
extension for the mode summation with respect to the Matsubara mode compared with 
that of the scalar. We have found that there is no nonanalytic term in the effective potential for the case of the fermion.
The result for the case of the fermion has immediately led to the conclusion that there also exists no nonanalytic 
term for the case of the scalar field satisfying the antiperiodic boundary condition for at least one spatial $S_i^1$ direction.

We have obtained some insight on the nonanalytic terms in the effective potential.
We have found that the nonanalytic term can appear only when there exists the zero mode associated with each of all the $S^1$'s
in the Matsubara and Kaluza-Klein modes. This observation may explain that the nonanalytic term depends on all the scales 
$L_0, L_1,\cdots, L_p$ like in Eq.(\ref{4.17}) (or Eq.(\ref{4.24})) and further that the effective potential 
has no nonanalytic term for the fermion due to the 
lack of the zero mode for the Euclidean time direction irrespective of the absence or presence of the zero mode
for the spatial $S_i^1~(i=1,2,\cdots, p)$ direction.

%%%%%% HERE COMES THE DISCUSSIONS ON GAUGE FIELD %%%%%%%%%%
Equipped with the result obtained for the cases of the scalar, we can also mention about the nonanalytic terms for 
the case of a higher dimensional gauge field on $S_{\tau}^1\times R^{D-(p+1)}\times\prod_i^p S^1_i$. 
Let us consider the $D$-dimensional gauge 
field $A_M(\tau, x^k, y^i)$, whose component gauge fields are written as 
\begin{align}
A_M(\tau, x^k,y^i)=\Bigl(A_{\tau}(\tau, x^k,y^i), A_k(\tau, x^k,y^i), A_i(\tau, x^k,y^i)\Bigr).
\label{7.1}
\end{align}
%%%%%%%%%%%%%%%%
We need to specify the boundary condition for the Euclidean time direction and the spatial
$S_i^1$ direction. The boundary condition for the Euclidean time direction must be periodic, {\it i.e.}
\begin{align}
A_M(\tau+L_0,x^k,y^i)=+A_M(\tau,x^k,y^i)
\label{7.2}
\end{align}
because of the quantum statistics. One can choose the boundary conditions 
of the $A_k(\tau, x^k, y^i)$ and the $A_i(\tau, x^k, y^i)$ for the spatial $S_i^1~(i=1,\cdots, p)$ direction 
%
%
%\begin{align}
%A_k(\tau, x^k, y^i+L_i)=\e^{2\pi i\eta_i}A_k(\tau, x^k, y^i),\quad
%A_i(\tau, x^k, y^i+L_i)=\e^{2\pi i\eta_i}A_k(\tau, x^k, y^i)
%\end{align}
%
%
to be $0$ or twisted under the assumption that the Lagrangian density
must be single valued. The gauge field $A_M(\tau, x^k, y^i)$ can be massive through the Higgs
mechanism and it has the field-dependent mass such as $M(\varphi)$. If we restrict the twisted boundary condition
for the spatial $S_i^1$ direction to the antiperiodic 
boundary condition, the nonanalytic term in the effective potential for each 
gauge field in Eq.(\ref{7.1}) is reduced to the cases of the scalar field studied in 
the sections $4$ and $6$. Thus, the result has already been 
obtained\footnote{If the component gauge field $A_i$ has the zero mode, it 
can acquire the vacuum expectation value (VEV) through the dynamics of the Wilson line phase\cite{hosotani}. The 
VEV is removed by the field redefinition to twist the boundary condition for the spatial direction
of the matter field.}.

%
%
%
%
%
%%%%%%% Future %%%%%%%%%%
%We have studied the nonanalytic term, paying attention to the type that cannot be written in the form of any 
%positive integer power of field-dependent mass squared, in the effective potential at finite temperature. 
%
%
%
In addition to the nonanalytic terms we have studied, there is another type of the nonanalytic 
term, $\log M$ in the effective potential, as mentioned 
in the sectionse $4$. Such the term arises from the double pole of the integrand, for 
example, in the residue integration of Eq.(\ref{4.9}). In order to understand the whole nonanalytic structure of the effective 
potential with respect to $M$, one has to study such a term extensively as well. 
%
%
%
%
%
%For the case of the scalar field satisfying the periodic boundary condition for 
%the spatial $S_i^1 (i=1,\cdots, p)$ direction, there exists only one nonanalytic terms (4.16) (or (4.23)) in the effective potential 
%in one-loop approximation for $D-(p+1)=$ odd. 
%
Our analyses have been carried out at the level of one-loop approximation for the effective potential at finite temperature.
One may wonder what type of the nonanalytic terms besides the one obtained in the paper can emerge beyond the one-loop calculation.
In connection with higher loop calculations, we are also interested in the behavior of the $\log M$ term, which actually 
stands for genuine quantum effects. These are under investigation and will be reported elsewhere.

%
%
%
%
%
%
%
%
%\newpage\noindent
%%%%%%%%%%%%%%%%%%%%%%%%%%%%%%%%%%%%%%

%%%%%%%%%%%%%%%%%%
\begin{center}
{\bf Acknowledgement}
\end{center}
%%%%%%%%%%%%%%%%%%%%
This work is supported in part by Grants-in-Aid for Scientific 
Research [No.~18K03649 and No.~23K03416(M.S.)] from the Ministry of 
Education, Culture, Sports, Science and Technology (MEXT) in Japan.
%%%%%%%%%%%%%%%%%%%%%%%
%
% Here comes the appendix
%%%%%%%%%%%%%%%%%%%%%%%%
%\appendix
%
%
%\section{Comments on the Dolan-Jackiw's calculation}
%Let us comment on the Dolan-Jackiw's calculation\cite{dj}.
%

%
%
%
%
%
%
%
%
%
%%%%%%%%%%%%%%%%%%%%%%%%%%%

\end{document}